\documentclass[fleqn,twoside]{article}
\usepackage{espcrc2}

\input{psfig}

\newcommand{\AmS}{{\protect\the\textfont2
  A\kern-.1667em\lower.5ex\hbox{M}\kern-.125emS}}

\title{The high energy view of blazars}

\author{Gabriele Ghisellini
\address[OAB]{Osservatorio Astronomico di Brera,
        Via Bianchi 46, I--23807 Merate, Italy}
}
       
\begin{document}

\begin{abstract}
{\it 
Beppo}SAX contributed substantially to our understanding
of the physics of blazars.
This has been made possible mainly by its wide energy range
and especially by its high energy detector.
Together with the information coming from still higher energies
we know at last the entire spectral energy distribution (SED) 
of blazars.
Different subclasses of blazars have different SEDs, 
which seem to form a sequence, whose main parameter is 
the bolometric luminosity.
Physically, the blazar sequence can be the result of different cooling 
rates by those electrons emitting most of the radiation we see.
Blazars are among the most active sources we know of, and the 
coordinated variability at high energies gives very important
clues about the physics of their jets: initially, they must transport 
energy in a dissipationless way.
They radiate most of the entire bolometric luminosity we see
at a preferred distance of a few hundreds of Schwarzchild radii.
For powerful blazars, the emitted radiation must be a small 
fraction of the power transported by the jet, since the bulk 
of it is required to energize the radio lobes.
These are the jets which likely have bulk relativistic
speeds up to hundreds of kiloparsecs, as suggested by 
the bright X--ray knots detected by Chandra in many extended jets.

\vspace{1pc}
\end{abstract}

\maketitle

\section{Introduction}

One of the most important discoveries of recent years is that
almost all galaxies contain a black hole in their centers,
with a mass approximately $10^{-3}$ that of the bulge.
In fact the correlations found between the black hole mass and either
the optical luminosity of the bulge 
(e.g. Kormendy \& Richstone 1995; Magorrian et al. 1998),
or the star dispersion velocity
(e.g. Ferrarese \& Merrit 2000; Gebhardt et al. 2000)
allow to measure the black hole mass with some confidence, and
therefore to measure the activity of the accretion disk (if at all 
present) in units of the Eddington luminosity.
For reasons we do not understand yet, 99 per cent of these black holes
are silent, in the sense that the radiation produced by the associated
accretion is null or less than the radiation produced by the stars.
Radio--quiet active galactic nuclei are the remaining 1 per cent,
and roughly 10 per cent of the latter are the radio--loud jetted
objects.
We now believe that all jets, or a significant fraction of them,
contain plasma moving at relativistic speeds corresponding
to bulk Lorentz factors $\Gamma$ around 10, producing radiation
which is seen amplified by observer in the ``beaming cone" $1/\Gamma$.
Thus these objects appear the most luminous and active.
We call them blazars.

\begin{table} 
\caption{{\it Beppo}SAX blazars}
\begin{tabular}{lrrr}
\hline
               &Total &FSRQs &BL Lacs  \\
\hline
Observations   &177   &54    &123  \\
Sources        &89    &29    &60  \\
P.I.           &24    &12    &18  \\
PDS Detect.    &47    &19    &28  \\
\hline
\end{tabular}
\label{table1}
\vskip -0.5 true cm
\end{table}

\begin{figure}
{\hskip -1.2 true cm \psfig{figure=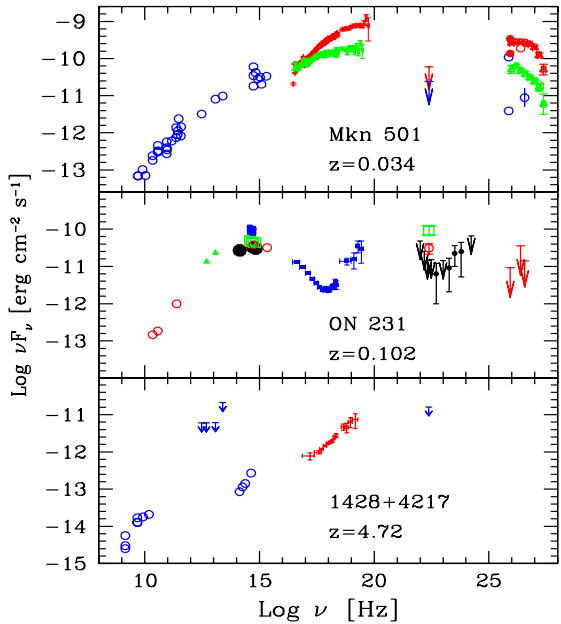,width=10cm,height=9cm}}
\vskip -1 true cm
\caption{
Examples of SED of blazars for which {\it Beppo}SAX  
observations were crucial. Adapted from Pian et. al. (1998),
Tagliaferri et al. 2001 and Fabian et al. (2001)
}
\label{sax}
\end{figure}

Soon after the launch of the Compton Gamma Ray Observatory
({\it CGRO}) 
in 1991, it was discovered that blazars, as a class, are very strong
$\gamma$--ray emitters above 100 MeV (Hartman et al. 1999).
Furthermore, approximately in the same years, ground based Cherenkov 
telescopes found that low power BL Lacs are also very strong 
TeV emitters (see e.g. Catanese \& Weekes 1999).
These discoveries were complemented by X--ray observations,
and in this field {\it Beppo}SAX played a major role, thanks
especially to its PDS instrument, sensitive between 15 and 100 keV,
which detected many blazars of all kinds (see Tab. \ref{table1}).
The {\it Beppo}SAX observations, which in many cases detected
blazars in the entire three decades energy range [0.1--100 keV], helped to
construct for the first time a meaningful spectral energy distribution 
(SED) for blazars, allowing to see unifying trends and sequences.
Fig. \ref{sax} shows, as illustration, the SED of three blazars
(in order of luminosity): {\it Beppo}SAX was crucial for all three,
establishing an unprecedented behavior for Mkn 501; introducing the concept
of ``intermediate blazars" through the X--ray spectral shape of sources
like ON 231; and discovering a very hard and powerful X--ray emission
in high redshift blazars even undetected by EGRET, such as 1428+4217 
(with $z$=4.72 it is the most distant known radio--loud quasar).
In addition, we could also study the variability properties in each band and 
between different bands, and revive theoretical studies of relativistic jets.

\section{The $\gamma$--ray zone}

Observations of blazars at high energies showed a strong
and (almost always) correlated variability of fluxes at the 
peaks of the SED.
The first consequence of these results was to abandon the idea
of a smooth inhomogeneous jet producing high frequencies
at its base and lower frequencies in its outer part:
most of the radiation we see should instead be produced
in a single zone of the jet
\footnote{Excluding the radio emission, which is self--absorbed
at these jet scales and should come from more extended
regions of the jet.}.
In the following subsection I review the basic arguments
leading to pinpoint the dimensions and the distance from the
black hole of this ``$\gamma$--ray zone".

\subsection{Low entropy inner jets}

The very fact that blazars are strong $\gamma$--ray emitters
implies that the produced $\gamma$--rays are not absorbed, 
and this in turn implies that: 
\begin{enumerate}
\item the emitting source cannot be too compact, to avoid
absorption of $\gamma$--rays through photon--photon collisions
within the source;
\item the $\gamma$--ray emitting source cannot be too
close to important sources of X--ray photons, such as a hot
accretion disk corona, to avoid absorption of $\gamma$--rays
with photons produced externally to the emitting region.
\end{enumerate}
Point 1) leads to the requirement of bulk relativistic motion, 
since the observed compactnesses are large (i.e. huge luminosities 
and small dimensions as indicated by the short variability
timescales, see e.g. Dondi \& Ghisellini 1995).
Point 2) leads to the requirement that the $\gamma$--rays
we see are produced beyond a critical distance from the 
black hole and the accretion disk, which is likely to produce
X--rays through its hot corona (Ghisellini \& Madau 1996).
Blandford (1993) and Blandford \& Levinson (1995)
introduced the concept of ``$\gamma$--ray photosphere",
proposing a jet which dissipates energy also very close to
the accretion disk.
In this model the $\gamma$--rays produced within the
$\gamma$--ray photosphere are absorbed and create pairs,
loading the jet.

\begin{figure}
\vskip -0.5 true cm
{\hskip -0.8 true cm \psfig{figure=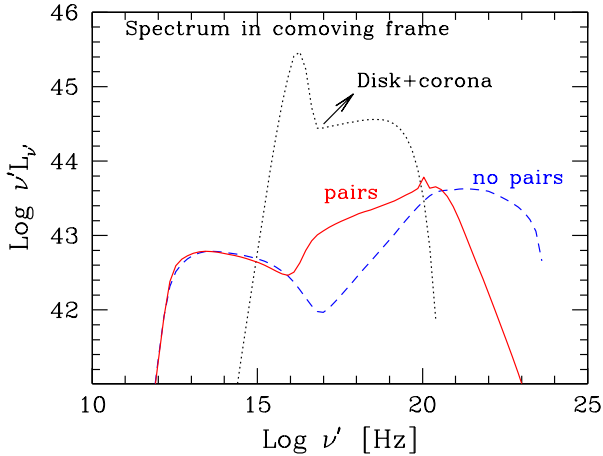,width=9cm,height=9cm}}
\vskip -3 true cm
\caption{
Synchrotron and Inverse Compton spectrum as observed
in the frame comoving with the emitting blob,
of radius $R=10^{15}$ cm moving with $\Gamma=15$.
The blob is $10^{16}$ cm away from an accretion disk
assumed to emit $\sim 10^{46}$ erg s$^{-1}$.
Note how the pair cascade process significantly
softens the spectrum, especially at X--ray frequencies.
}
\label{blandford}
\end{figure}

In this case the jet is dissipative from its start.
Ghisellini \& Madau (1996) instead argued that the
accretion disk, besides providing X--rays which
interact with the $\gamma$--rays produced within the $\gamma$--ray 
photosphere, is producing UV radiation, which cools the just
born relativistic pairs through the Inverse Compton process.
In this case the pairs emit almost all their energy in (beamed)
X--ray radiation, with a luminosity comparable to the 
observed $\gamma$--ray power.
Since this is not observed, this process cannot occur:
the jet therefore cannot dissipate much of its kinetic
energy at small distances from the accretion disk.
Fig. \ref{blandford} illustrates this point.
For simplicity, the optical-UV emission produced by the
accretion disk is approximated with a blackbody, 
while the X--ray flux produced by the corona is
assumed to be a power law (of energy index $\alpha_x=0.9$)
ending with an exponential cut
(dotted line in Fig. \ref{blandford}: the power
is assumed to be calculated by an observer comoving with a
blob moving with $\Gamma=15$).
The blob is assumed to be spherical, with a size $R=10^{15}$ cm
at a distance of $z=10^{16}$ cm from the accretion disk
(whose relevant size is assumed to be $R_{\rm disk}\sim 3\times 10^{15}$ cm).
The particle distribution responsible for the emission is
calculated through the continuity equation, 
assuming continuous injection of elecrons
distributed as a power law in energy between $\gamma_1=100$
and $\gamma_2=5\times 10^3$.
We can see the effects of neglecting (dashed line) or including
(solid line) pair production and pair reprocessing.
All the energy absorbed in the $\gamma$--ray band is re--emitted,
especially at X--ray energies.
Observations show instead that in powerful blazars the flux
in the X--ray range [2--10 keV] is particularly hard in shape and much
much fainter than the $\gamma$--ray flux.
We therefore conclude that in the presence of an accretion disk
(producing cooling UV radiation) and its corona (producing
X--ray photons, which are targets for the $\gamma$--$\gamma \to e^\pm$
process), the jet must transport its energy to a few hundreds of
Schwarzchild radii before radiating.

On the other hand, the short variability timescales we observe,
especially in the X--ray and $\gamma$--ray bands, constrain
the size of the emitting region to be of the order of
$\sim$1 light--day $\times\delta \sim 10^{16}$ cm or less,
where $\delta$ is the Doppler factor.
In a conical jet with aperture angle $\sim 0.1$, this
dimension corresponds to a distance $z\sim 10^{17}$ cm from the black hole.
We therefore conclude that there is a {\it preferred distance}
where most of the dissipation is taking place.

\subsection{No electron--positron pair cascades}

\begin{figure}
\vskip -0.6 true cm
{\hskip -0.8 true cm \psfig{figure=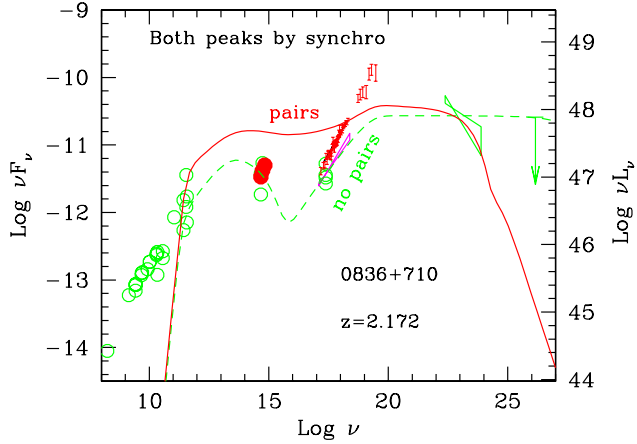,width=9cm,height=9cm}}
\vskip -3.2 true cm
\caption{
Pair cascades and the associated
pair reprocessing can greatly affect the SED of a blazars.
The reprocessing tends
to fill in the ``valley" between the two peaks of the SED.
In this example, the spectrum is due to two different 
electron populations, with different
high energy cut--offs and both peaks are due to synchrotron.
The magnetic field is required to be large
to avoid too much IC radiation (in this example $B=50$ G). 
If the high energy electrons were created through a pair--cascade
of ultra--high energy photons, then too many X--rays would be produced.
}
\label{proton}
\end{figure}

The argument about the required absence of pair
creation and pair reprocessing is more general
than discussed above.
In fact, also those models which invoke the existence of ultra--high 
energy protons, making pairs through interactions
with optical--UV photons, face the problem of how to avoid
pair cascades with the production of too many X--rays
(see also Sikora \& Madejski 2001).
As an example, Fig. \ref{proton} shows the effects
of neglecting/including the effects of pairs.
For this model, I have assumed two different electron injection
functions, one at low energies, responsible for the first
peak of the SED through the synchrotron process, and another population 
of electrons at much higher energies, responsible for the second 
peak of the SED through, again, the synchrotron process. 
This second electron population should correspond
to those leptons created through the proton--photon interactions,
as envisaged in the ``proton blazar" model by Mannehim (1993).

For illustration, the resulting spectra are compared to the SED
of 0836+710, one of the most distant blazars detected by EGRET,
with $z=2.17$. 
This source has a steep $\gamma$--ray spectrum, which {\it cannot}
be the result of $\gamma$--absorption, since otherwise the
created pairs would overproduce the X--ray flux even if any
external radiation is absent and even if the dominant radiation
mechanism is synchrotron, i.e. for magnetically dominated jets.

\section{The blazar sequence extended to low power BL Lacs}

\begin{figure}
\vskip -0.5 true cm
{\hskip -0.3 true cm \psfig{figure=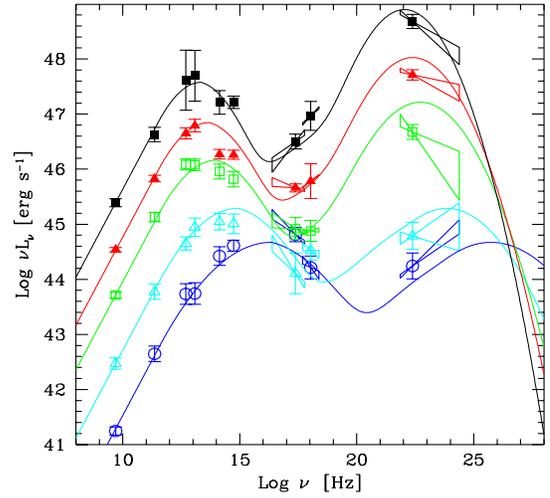,width=8.5cm,height=8cm}}
\vskip -1.8 true cm
\caption{The blazar sequence. From Fossati et al. (1998) as modified
by Donato et al. (2001).
}
\label{sequence}
\end{figure}
\begin{figure}
\vskip -0.6 true cm
{\hskip -0.8 true cm \psfig{figure=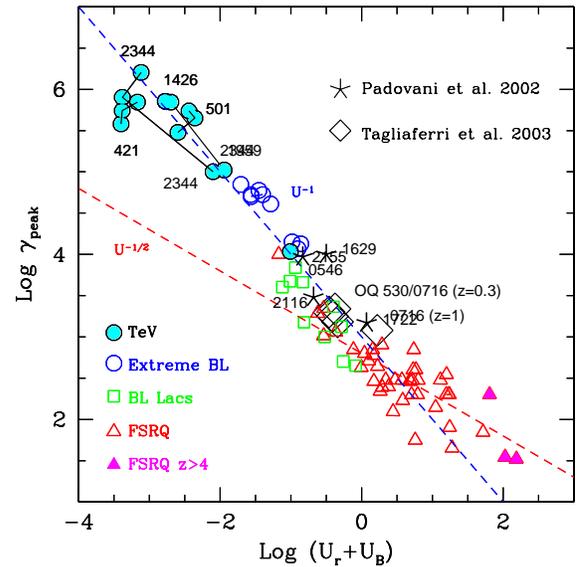,width=9cm,height=9.cm}}
\vskip -1.5 true cm
\caption{
The random Lorentz factor of the electrons emitting at the peaks of
the SED, $\gamma_{\rm peak}$, as a function of the comoving energy density
(radiative plus magnetic).
The dashed lines correspond to $\gamma_{\rm peak} \propto U^{-1/2}$ and
$\propto U^{-1}$ (they are not best fits). From Ghisellini et al. (2002), 
updated with the data fom Padovani et al. (2002) and Tagliaferri et al. (2003).
}
\label{gb_u}
\end{figure}

Fossati et al. (1998), collecting data from three complete sample
of blazars, demonstrated that the SED is controlled by the bolometric
observed luminosity, with both peaks shifting at smaller frequences when 
increasing the luminosity (see Fig. \ref{sequence}).
Furthermore, the dominance of the high energy peak increases when
increasing the bolometric luminosity.
This latter inference, however, was based on those few 
low power BL Lacs detected by EGRET.
The new data coming from Cherenkov telescopes suggest instead
that, in some cases, also in low power BL Lacs the high energy
peak dominates the bolometric output, especially considering
that the TeV emission could be severely absorbed by the diffuse
IR background through the $\gamma$--$\gamma$ process 
(see e.g. Stecker \& De Jager, 1997; for recent results on TeV sources
see Aharonian et al. 2002, Costamante et al. 2003).
This ``blazar sequence" can be explained by a different degree of
radiative cooling: in powerful blazars electrons cool faster,
producing a break in the electron distribution functions
at smaller and smaller energies when increasing the total
(radiation plus magnetic) energy density in the comoving frame.

We (Ghisellini et al. 2002) extended the 
blazar sequence to include ``extreme" BL Lacs, namely those low
power BL Lacs which are the best candidate to be TeV emitters.
Fig. \ref{gb_u} shows the random Lorentz factor emitting at the peaks
of the SED as a function of the energy density as seen in the 
comoving frame.
There are two branches: 
for ``extreme" BL Lacs we have 
$\gamma_{\rm peak}\propto U^{-1}$, while for the other blazars
$\gamma_{\rm peak}\propto U^{-1/2}$.
Note that these behaviors corresponds, respectively, to a
constant cooling time at $\gamma_{\rm peak}$ 
[i.e. $t_{\rm cool}(\gamma_{\rm peak}) \propto (\gamma_{\rm peak}U)^{-1}$=const],
or at a constant cooling rate at $\gamma_{\rm peak}$ 
[i.e. $\dot\gamma(\gamma_{\rm peak})\propto \gamma_{\rm peak}^2U =$ const].

We have interpreted this behavior by assuming that in all blazars
the acceleration mechanism injects relativistic electrons 
between $\gamma_1$ and $\gamma_2$, but only for 
a finite time, which is of the order of the light 
crossing time of the source, $R/c$.
In a powerful blazars all electrons cool in a timescale shorter 
than $R/c$, and the resulting particle distribution, after this time, 
has a break at $\gamma_{\rm peak}=\gamma_1$.

In low power BL Lacs, instead, only the very high energy particles
can cool in $R/c$, and after this time the particle distribution
will have a ``cooling" break at some $\gamma_{\rm c}$
(with $\gamma_{\rm c}>\gamma_1$) which is
the energy where the cooling time is equal to $R/c$, 
yielding $\gamma_{\rm peak} =\gamma_{\rm c} \propto U^{-1}$.

\subsection{Variability above and below the peaks}

A ``snapshot" taken at the end of the injection 
time would catch a flare at its maximum.
The assumption that the injection of particles lasts only for the finite
time $R/c$ implies that the source would always vary on the
timescale $\sim R/c$ (shortened by relativist effects) above the
``cooling frequencies" $\nu_{\rm c}$ 
(namely the synchrotron and the inverse Compton
frequencies produced by electrons with $\gamma>\gamma_{\rm c}$),
while the flux below $\nu_{\rm c}$ should vary with the cooling timescale
(i.e. with the characteristic $t_{\rm cool} \propto \nu^{-1/2}$
behavior).

This translates in a simple prediction: in low power BL Lacs,
in which the peaks correspond to $\nu_{\rm c}$, 
we should see a characteristic variability behavior: 
at frequencies above the peaks of the SED the flux should vary
with the same timescale, indicative of the size of the emission region,
while at frequencies below the peaks the flux should vary 
with a timescale given by 
$ct_{\rm var}\delta\sim (R/c)(\nu_{\rm peak}/\nu)^{1/2}$,
where $\delta$ is the Doppler factor. 

In powerful blazars, instead, the peaks are determined by the injection
energy $\gamma_1$, and not by $\gamma_{\rm c}$. 
This means that at frequencies both above and below the peak
the flux should vary on the same timescale given by $c t_{\rm var} \delta 
\sim R/c$
\footnote{
These arguments strictly apply for a one--zone model.
In reality, it is very likely that, at any given time, we are
observing a number of emitting regions.
This is even more likely at smaller (radio and IR) frequencies.
However, the above considerations can be relevant when a single region
dominates the emission, i.e. during strong (and short) flares.
}.

\section{Jet power}

We believe that powerful blazars are associated with FR II galaxies,
with prominent radio lobes.
We can use equipartition arguments to find a minimum energy
content in these structures, and by dividing it by the source lifetime
we have an average power that the lobes require to exist.
This is then a lower limit to the power of jets (Rawlings \& Sanders 1991).
This power is greater than the power radiated by the jet along its
way, implying a small (i.e. 10 per cent) efficiency in converting 
bulk kinetic energy (or Poynting flux) into radiation.
The jet energy carriers could be electrons and protons,
electron--positrons pairs, or Poynting flux.
Recently, the circular polarization discovered in the jet of 3C 279
by Wardle et al. (1998) led these authors to suggest that pairs 
are dominant, but later work (Ruszkowski \& Begelman 2002) demonstrated
that this is not necessarily the case.
Another uncertainty is if the power of jets changes along the way,
which is unlikely in powerful blazars and FR II, 
but perhaps possible in low power BL Lacs and FR I.
It is therefore interesting to measure the power of jets
at different scales.
Up to now there are three jet scales where an estimate of the
jet power is possible:

\begin{figure}
\vskip -0.1 true cm
{\hskip -0.3 true cm \psfig{figure=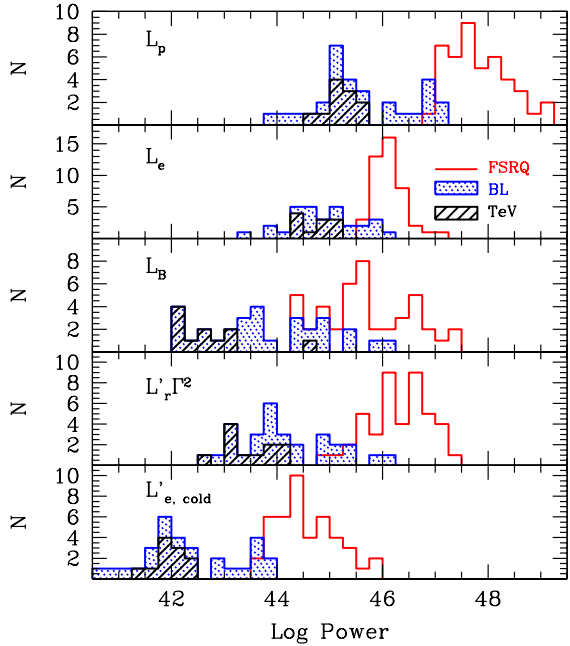,width=9cm,height=9cm}}
\vskip -1 true cm
\caption{Distribution of jet powers
for the blazars considered in Celotti \& Ghisellini (2003).
$L_{\rm p}$ is the power carried by protons assuming one
proton per emitting electron;
$L_{\rm e}$ is the power carried by the emitting electrons;
$L_{\rm B}$ is the Poynting flux;
$L^\prime_{\rm r}\Gamma^2$ is the power radiated by the jet
and $L_{\rm e, cold}$ is the power carried in the form of 
electron rest mass (without including the relativistic random
energy of the particles).
}
\label{power}
\end{figure}

\begin{enumerate}
\item
the $\gamma$--ray zone ($\sim$0.1 pc): 
this is the most dissipative zone, and
we can measure the number of particles, the magnetic field, the
size and the bulk Lorentz factor through the modeling of the SED
(Ghisellini 2001;  Ghisellini \& Celotti 2002;  Maraschi \& Tavecchio 2003;
Celotti \& Ghisellini 2003);
\item
the VLBI scale ($\sim$1--10 pc), where we can measure directly the apparent speed
and the size (Celotti \& Fabian 1993; Celotti et al. 1997); 
\item
the large scale jet ($\sim 10^5$ pc) as recently observed by Chandra in X--rays
(Celotti et al. 2001, Tavecchio et al. 2000, Ghisellini \& Celotti 2001a).
\end{enumerate}

Fig. \ref{power} shows the results obtained for the $\gamma$--ray zone
by Celotti \& Ghisellini (2003), considering all blazars
with sufficient data to constrain the adopted model, which is a single
zone synchrotron inverse Compton model with finite injection time.
Due to the limited sensitivity of EGRET and Cherenkov telescopes,
this implies that the SEDs we have modeled correspond to 
a flare state of the sources.
Therefore the powers calculated in this way are probably upper values:
to estimate the average powers one should know the duty 
cycle (i.e. the fraction of the time spent during flares).
This will be possible with GLAST.
From Fig. \ref{power} we note:
\begin{itemize}
\item
The estimated jet powers are large, and often exceed
the power radiated by accretion (which can be estimated, in this beamed
sources, through the luminosity of the emission lines).
\item 
For powerful blazars (i.e. FSRQs sources) the power that the jet spends 
to produce radiation is in some case larger than the power carried 
in relativistic electrons responsible for the emission. 
\item
Also the Poynting flux is often less then the radiated power.
This is to be expected, since sources with a Compton flux dominating over the
synchrotron flux cannot have large magnetic fields.
\item
If we assume that there is a proton for each electron, then the 
power carried by protons is a factor 10--30 larger than the
radiated power, implying efficiencies of 3--10 per cent.
In this case the remaining power can reach and energize the radio lobes.
\item 
For low power BL Lacs, instead, there is not much difference between 
the power in electrons and in protons, since in these sources
the average random Lorentz factors of the electrons is $\sim 1000$.
\end{itemize}
We then conclude that, at least in powerful blazars, the proton
component of the jet must be energetically dominant (only 2 or 3
pairs per proton are allowed), unless the magnetic field present
in the emitting region is only a fraction of what the jet transports.

\section{Chandra jets}

Chandra detected X--ray emission from knots in jets at 
distances of tens to hundreds kpc from the nucleus, both 
in powerful flat spectrum radio sources whose jet is 
probably aligned with the line of sight and in radio 
galaxies whose jets are instead observed at large viewing angles 
(see Tavecchio et al. 2003 and references therein).

For aligned jets the most popular interpretation of the
observed X--ray emission is inverse Compton  
off the cosmic microwave background (CMB).
This interpretation has also the virtue to minimize 
the energy requirements (Celotti et al. 2001; 
Tavecchio et al. 2000; Ghisellini \& Celotti 2001a).
This however requires the jets to be highly relativistic even at the
largest scales: in the case of PKS 0637--712 the bulk Lorentz factor
$\Gamma$ should still be 10--15, a few hundreds kpc away from the
core.  
Bulk relativistic motion in fact implies that the CMB energy
density is seen boosted by a factor $\sim \Gamma^2$ in the plasma
comoving frame, and therefore can dominate over the local synchrotron
and magnetic energy densities.

If this interpretation is correct, it also implies that
the electrons producing the X--rays have random Lorentz
factors of the order of one hundred with radiative
cooling time much longer than the light crossing time.
The optical and radio radiation, instead, is believed
to be produced by synchrotron, by much more
energetic electrons, with cooling times (for the optical)
nicely coincident with the size of the knots, as measured by HST.
Available observations then pose an interesting problem: 
outside the bright knots the X--ray flux is dimming as fast
as the optical and radio fluxes, despite of the fact that it should be
produced by low energy electrons, which do not cool radiatively. 
Do they cool by adiabatic losses?
Is so, we then require several ``doubling radii" to 
``switch off" the X--ray emission.
The conclusion then is that the knot cannot be homogeneously filled
with the emitting particles, but must be composed by smaller sub--units.
In other words, the knot must be clumped: particles injected
in each clump can then lose energy by adiabatic losses
before reaching the borders 
of the knot, and produce a negligible amount of X--rays outside it
(Tavecchio et al. 2003).

Note also that even if the electrons have lost their random energy
once the reach the border of the knot, they still have
bulk motion energy, and they continue to scatter CMB
radiation.
Such ``large scale bulk Compton process" produces 
beamed radiation 
at frequencies $\nu\sim 3\times 10^{11} \Gamma^2$ Hz
(the redshift dependence drops out), potentially detectable by ALMA.
This could be a powerful test for the idea that these jets
are relativistic up to large scales.

\section{Internal shocks?}

The result on the jet power and on the spectral modeling of blazars mentioned 
above can have a satisfactory explanation in the internal shock scenario, 
in which the central engine works intermittently producing blobs moving 
at slightly different velocities and therefore colliding at some distance 
from the black hole, transforming a few per cent of the bulk kinetic 
energy in radiation (see e.g. Ghisellini 1999; Spada et al., 2001).
In fact this scenario easily explains:
\begin{itemize}
\item
The fact that most of the dissipation occurs at $\sim$hundreds of 
Schwarzchild radii, which is the distance of the first collisions 
between consecutive shells;
\item 
The observed variability, which is a built--in feature 
(but requires that the central engine works intermittently); 
\item
The fact that we need particle acceleration lasting for a finite time,
to explain the $\gamma_{\rm peak}-U$ correlation, since we have 
acceleration of particles for about one shell light crossing time;
\item
The fact that the jet produces radiation all along the way,
but with a reduced efficiency.
In fact, at large distances from the jet apex,
shells that have already collided are moving 
with Lorentz factors which are intermediate of the original ones. 
These shell will collide further, but with a reduced contrast between
their Lorentz factors, and therefore with a reduced efficiency.
\item
The blazar sequence, since the first energetic shell--shell collisions 
occur within the broad line region in powerful blazars, and outside
the BLR in BL Lacs (if we assume that the BLR is located at 
a distance which correlates with the accretion disk luminosity, 
see e.g. Kaspi et al. 2000).
In weak BL Lacs shell--shell collisions occur outside the BLR,
in a much less dense external photon environment, implying less severe cooling
(i.e. large $\gamma_{\rm peak}$) and a relatively more important SSC emission.
Intermediate cases should exist, where the first collisions occurs
sometimes within and sometimes outside the BLR. 
These sources should be characterized by a dramatic 
variability at high energies, such as observed in BL Lac itself 
(Bloom et al. 1997; Ravasio et al. 2002).

\end{itemize}

It is also a relatively simple scenario, allowing quantitative analysis
(see Tanihata et al. 2002 for an application to Mkn 421).
Besides all that, part of the appeal of this scenario lies on the possibility
that $all$ relativistic jets work the same way, therefore including gamma ray
bursts and galactic superluminal sources.

On the other hand, Blandford (2003) and  Lyutikov \& Blandford (2003) have 
proposed a purely electromagnetic jet, magnetically dominated.
This contrasts with our distribution of Poynting fluxes
for blazars, shown in Fig. \ref{power},
but this could be the results of having localized (clumped)
emission regions where the magnetic field is less than the
average (because of reconnection?).
We need thus a way to distinguish between the two scenarios,
bearing in mind that also in the internal shock scenario an acceleration
mechanism is required, which might use magnetic forces. 
However, the matter has to achieve its final bulk Lorentz factor quite 
rapidly, before the $\gamma$--ray dissipation zone.
One possibility to test the presence of a matter
dominated jet might be to find a feature in the X--ray spectrum
resulting from bulk Comptonization occurring
at the base of the jet, as discussed by
Sikora et al. (1997) and Sikora \& Madejski (2003). 
In this case some excess emission is expected, at a 
frequency $\nu_{\rm disk}\Gamma^2\sim$1 keV, where $\nu_{\rm disk}$ is the peak frequency of the disk radiation.

\section{Parent populations}

It is commonly believed that BL Lac objects are aligned FR I 
sources, while the more powerful FSRQs are aligned FR II sources.
This idea, originally put forward by Blandford \& Rees (1978),
has been tested by Urry and Padovani (see Urry \& Padovani 1995
and references therein).
Note that there can be ``classical" BL Lacs
(for instance: PKS 0537--441) which have line equivalenth widths
less than the canonical 5 \AA ~ value, but which have very luminous
lines nevertheless: these could well be FSRQs with a particularly enhanced
non--thermal continuum, and then belonging to the FR II class.

\begin{figure}
{\hskip -0.8 true cm \psfig{figure=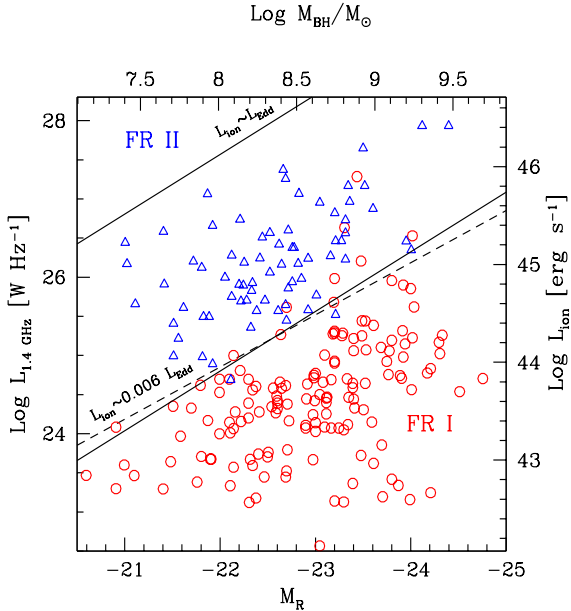,width=8.5cm,height=9cm}}
\vskip -1.5 true cm
\caption{The radio jet power -- host optical magnitude plane with the line 
sharply dividing FR~I from FR~II (dashed line, from  Ledlow \& Owen 1996). 
According to reasonably well established correlations this plane 
is equivalent to an accretion  power vs black hole  mass plane 
(right and upper axis). 
The central diagonal line represents $L_{\rm ion} \sim 6\times 10^{-3} 
L_{\rm Edd}$. From Ghisellini \& Celotti (2002).
}
\label{fr12}
\end{figure}

The recent possibility to determine the mass
of the black hole through observations of the host galaxy
made possible to investigate the black hole masses of both FR I and
FR II, to see if they are different.
We (Ghisellini \& Celotti 2001b) considered the ``Ledlow \& Owen" (1996) 
plot, in which the two types of radio galaxies are neatly divided
by a line in plane of radio luminosity vs optical luminosity of the
host galaxy.
In fact the optical luminosity of the host allows to
estimate the black hole mass, while the radio luminosity
correlates with the accretion disk luminosity.
Fig. \ref{fr12} shows the result: the dividing line
between FR I and FR II is indistinguishable from the line
corresponding to an accretion luminosity at the 0.6 per cent
of the Eddington one.
This bears the intriguing possibility that the diversity
of jets is due to a property of the very nucleus of the AGN, and
not to the different ambient into which the jet propagates.
Namely, it is the accretion mode that could be different,
since for ratios $L/L_{\rm Edd}<10^{-2}$ one expects 
that advection dominated accretion becomes important.

\section{Conclusions}

Extragalactic jets can have powers of the order of,
but even greater then, what radiated by accretion.
This is most clear in BL Lac objects, which lack
any sign of the accretion activity, but can become
dramatically evident if we consider gamma ray bursts (GRB)
as jetted sources.

We are starting to understand the phenomenology of 
extragalactic jets, even if the issue of their formation,
collimation and acceleration are still open.
It is possible that blazar jets are similar to the
jets of galactic superluminal sources and to the 
jets of GRBs. 
If true, it is very fruitful to study their similarities,
differences, and the scaling laws.
After all, the ``internal shock" scenario was invented in the AGN
field (Rees 1978), then it became the ``standard" model to explain the
prompt emission of GRBs, and it is again useful for blazars.

After the death of {\it Beppo}SAX and {\it CGRO} 
the observational prospects can become favorable again
to blazar studies both by satellites 
(SWIFT, ASTRO--E2, and especially AGILE and GLAST)
and by the new generation of ground based Cherenkov telescopes,
(H.E.S.S., MAGIC, CANGAROO III and VERITAS), which should be
a factor 10 more sensitive and have a smaller energy threshold, 
enabling them to detect distant (i.e $z>1$) sources.
Blazars will be observed in all high energy bands almost
without energy gaps, with very important consequences
especially on cosmology, through the measurements they will 
allow on the IR through the UV cosmic backgrounds.

\vskip 0.5 true cm

\noindent
{\bf Acknwoledgments} --- It is a pleasure to thank 
Annalisa Celotti and Fabrizio Tavecchio for continuous 
interactions and collaborations.

\end{document}